\documentclass{article}

\usepackage{arxiv}

\usepackage[utf8]{inputenc} 
\usepackage[T1]{fontenc}    
\usepackage{hyperref}       
\usepackage{url}            
\usepackage{booktabs}       
\usepackage{amsfonts}       
\usepackage{nicefrac}       
\usepackage{microtype}      
\usepackage{lipsum}

\title{A  MacDowell-Mansouri type formulation for Einstein manifolds.}

\author{
  J. E. Rosales-Quintero \\
  Preparatoria 2 de octubre de 1968, Universidad Aut\'onoma de Puebla,\\ P.O. Box
1364, 72000 Puebla, México\\
  \texttt{ jeduardo.rosales@correo.buap.mx, jose.rosalesquintero@viep.com.mx} \\
}

\begin{document}
\maketitle

\begin{abstract}
We study a pure connection formulation plus algebraic constraints  in four spacetime dimensions where the gauge group  $G \supset SO(1, 3)$.  We show that the action has, as particular cases, the MacDowell-Mansouri and the Stelle-West formulations for gravity. Also, under adequate specification of the constraint terms, we obtain Einstein manifolds, i.e., torsionless conformally flat manifolds.  

\end{abstract}


\section{Introduction}
Gravity, as it was  formulated by Einstein at the beginning of the last century, considers the metric as the fundamental object. The resulting field equations involve high non-linear second order derivatives. In order to attain a gravity formulation in
which its equations of motion consist in first order derivatives, an auxiliary field is introduced, the connection field. This field can be seen as the  analogous to a momentum variable that takes the Lagrangian formulation up  to the Hamiltonian formulation. But the introduction of the connection not only permits a simplification on the differential equations, it lets us construct principal fiber bundles over the spacetime, letting us reveal gravity as a gauge theory. But we have to take into account that even if the connection is introduced into the formulation, it usually takes a secondary role as an auxiliary object, some examples of such formulations are the Palatini formulation of gravity, BF Plebanski type formulation (real or complex case) and the Cartan formulation (see for example \cite{Peldan1994, plebanski, Krasnov2018}). These kind of formulations are pretty interesting by themselves, but in this work we will consider formulations of gravity where  the main, or fundamental, field is the connection. Examples of such formulations were given by Capovilla, Dell and Jacobson, where in the late of the 80's of the last century, they introduced a pure connection formulation for gravity (CDJ Action)  where the fundamental field is a complex $SU(2)$ connection \cite{Capovilla1991}, also, MacDowell and Mansouri (MM) \cite{MacDowell1977, Lopez-Dominguez2018} and Stelle and West \cite{Stelle1979, Stelle1980, DIaz-Saldana2021} in the middle of the 70's proposed independently pure connection actions for gravity  where, in both formulations, instead of enlarging  the spacetime,   they enlarged the gauge group $G$, such that $SO(1,3) \subset G$. In the last decade, Krasnov introduce a pure connection formulation for a diffeomorphism invariant $SU(2)$ gauge theory, but since the fundamental fields are complex, then reality conditions must be introduced  \cite{Krasnov2011}. More recently, Mitsou introduce a pure $SO(1, 3)$ connection action for full General Relativity (GR), where the connection is real definite and reality conditions are not needed at all \cite{Mitsou2019}. \\
On the other hand, a theory of gravity in four spacetime dimensions whose only solutions are gravitational instantons, i.e. Einstein  metrics with a half of the Weyl curvature vanishing \cite{Besse1987}, is known under the name of Self-dual gravity (SDG). Physically, SDG  is a theory that describes two propagating degrees of freedom, it is diffeomorphism  invariant
and it has a coupling constant of negative mass dimension, but even with this problem it give rise to a perfectly  well-behaved quantum theory, i.e. there are not quantum divergences and it is finite \cite{Krasnov2017}.\\
In this work, we will consider a pure connection formulation for gravity, that we will name it as $\Psi(F)$-formulation, introduced by the author himself in a pair of works for gravity and supergravity \cite{Rosales-Quintero2016, Rosales-Quintero2020} and inspired by the works of Torre \cite{Torre1990} and Smolin \cite{smolin2002quantum}. In  $\Psi(F)$-formulation we do not  consider, in principle, full GR,  instead this action  let us construct maximally symmetric spaces, i.e., Einstein torsionless conformally flat manifolds, so this formulation is a SDG formulation. But, as we shall later see, when we consider a higher dimensional gauge group, the MM and the SW formulations are contained into the $\Psi(F)$-formulation, this means that when we enlarge the gauge group, $\Psi(F)$-formulation can be rise to a full GR formulation. Finally, we have to mention that  this formulation makes use of an adequate gauge invariant constraint terms. \\  
The organization of this paper is as follows. In Sec. 2  We present a brief introduction to the $\Psi(F)$-formulation, together with an example for a gauge group $SO(1, 3)$ (for full details we refer you to the paper \cite{Rosales-Quintero2020} ). In  Section 3, We show that if we consider the  $\Psi(F)$-formulation for an enlarged group $SO(1, 4)$ we have as special cases the MM formulation and the SW formulation for gravity. Also, we introduce  the constraint terms that let us obtain explicitly  Einstein torsionless conformally flat manifolds. We conclude with a brief discussion and outlook.\\
In this paper, we have labeled $\mathfrak{so(1,4)}$ Lie algebra indices by capital Latin letters $\{ A, B, C, \ldots \}$,   $\mathfrak{so(1,3)}$ Lie algebra indices by the
beginning of the Latin alphabet lowercase letters $\{a, b, c, \ldots \}$ and Greek alphabet letters for spacetime indices $\{ \mu, \nu, \rho, \ldots \}$. Also, we consider the Minkowski metric as $\eta_{ab}=diag(-1, 1, 1, 1)$ and $\eta_{AB}=diag(-1, 1, 1, 1, 1)$, and we establish $\eta_{AB, CD}=\eta_{AC}\eta_{BD}-\eta_{AD}\eta_{BC}$ and $\eta_{ab,cd}=\eta_{ac}\eta_{bd}-\eta_{ad}\eta_{bc}$. In addition, we have defined the totally anti-symmetric tensor $\epsilon$ with $\epsilon^{01234} =+1$ and $\epsilon_{01234} =-1$ and,  in four dimensions,  $\epsilon^{0123} =+1$ and $\epsilon_{0123} =-1$. Finally, we establish $G_{(ab)} = G_{ab} + G_{ba}$ and $G_{[ab]} = G_{ab} - G_{ba}$.

\section{A pure connection formulation  for gravity}
 Since the connection one-form is the main ingredient in pure connection formulations for gravity, let us consider a principal $G$-bundle $P$ over $\mathcal{M}$, where $G$ is the gauge group, $\mathfrak{g}$ its Lie algebra and $\mathcal{M}$, the space-time,  is a 4-dimensional oriented smooth manifold. Let us consider $\mathcal{M}=\mathbb{R}\times \Sigma$, where $\Sigma$ is  a compact manifold  without boundary, and $\mathbb{R}$ represents an evolution parameter, in other words, $\mathcal{M}$ is globally hyperbolic, this is so, in order to avoid bad causality properties. The action that will be used in this work is given by
\begin{equation} \label{eq: Rosales Action}
    S[A, \Psi, \rho]=\int_{\mathcal{M}} Tr(\Psi(F)\wedge F+\rho\Psi).
\end{equation}
where $A$ is a $\mathfrak{g}$-valued connection one-form and $F$ is the two-form field strength constructed, as usual, from $A$ as $F=dA+A\wedge A$. Also, $\Psi$ is a Lagrange multiplier which behaves as a endomorphism when acts on the field strength $F$, otherwise it is a Lie algebra bivector valued field, and $\rho$ is a four-form that constraints the functional form of $\Psi$ \cite{Rosales-Quintero2016, Rosales-Quintero2020}. We have to note that no cosmological constant enters explicitly into the action, since it arrives in a different manner due to algebraic equations of motion \cite{Krasnov2018, Torre1990}.\\
As we can observe from (\ref{eq: Rosales Action}), in this kind of formulations,  the  gauge field for a corresponding symmetry group acts as the fundamental field. But in order to describe  gravity by action (\ref{eq: Rosales Action}), it is important to analyze the constraints involved by the Lagrange multipliers $\Psi$ and $\rho$. To this end, let us take into account (\ref{eq: Rosales Action}), and let us consider $\mathfrak{g}$  such  that its generators -labelled by $\{t_{p}\}$ where $p=1, \ldots $ dim(G)- accomplish
\begin{displaymath}
  [t_{p},t_{q}]=f_{pq}^{\ \ r}t_{r}
\end{displaymath}
where  $f_{pq}^{\ \ r}$ are the  structure constants for the commutator of two Lie algebra generators. Then, the Cartan-Killing form (CKf), is built, as usual, as $\kappa_{pq}=\frac{1}{I_{ad}} \ tr(t_{p}t_{q})$, where the appearance of $I_{ad}$ it is due that the  metric is defined upon a overall constant factor. Now, it is important to point out that we are defining two different traces: the first one is introduced into the action (\ref{eq: Rosales Action}) and into the definition for $\Psi(F)$, as we will notice further down in the text. The second one, is presented into the definition of the CKf. The trace "tr" is the usual trace, but the trace "Tr" is introduced here since for some Lie algebras there are more that one invariant trace that can be defined, and this trace, $Tr$, is built up from linear combination of the different invariant traces that can be considered into some Lie algebra. Let us suppose that in its Lie algebra $\mathfrak{g}$ of our gauge group under consideration $G$ there are $n$ different invariant traces that can be defined ($n \in \mathcal{N}$ ), then $Tr$ is defined as
\begin{equation}
    Tr(t_{p}t_{q})=a_{1} tr^{1}(t_{p}t_{q})+a_{2}tr^{2}(t_{p}t_{q})+\ldots+a_{n} tr^{n}(t_{p}t_{q})
\end{equation}
being the superscript index $n$ label each different invariant traces and all $a_{i}$ are scalars. In the next section we will present the case of $SO(1,3)$, it has two invariant traces that can be defined (see for example \cite{Rosales-Quintero2020} and the references therein). If we consider  a Lie algebra where we can defined a unique trace, such as  in the case of semisimple compact groups as $SU(2)$ or $SO(3)$,  then $Tr=tr$.\\
Now, let us go back to the action (\ref{eq: Rosales Action}), the connection one-form and the field strength are  given by $A=A^{r}t_{r}$ and $F=F^{p}t_{p}$ respectively. Now, as pointed out by \cite{Rosales-Quintero2020}, the endomorphism $\Psi$ acts on $F$ as follows:
    \begin{equation}
        \Psi(F)=\Psi^{pq} F^{r}\ Tr(t_{p}t_{r})t_{q}=\Psi^{pq}\mathfrak{F}_{p} t_{q}
    \end{equation}
where we have defined $ F^{r}\ Tr(t_{p}t_{r})=\mathfrak{F}_{p}$. Let us substitute back all this information into our initial action (\ref{eq: Rosales Action})
\begin{equation}  \label{eq: Rosales Action indices version}
    S= \int_{\mathcal{M}} \ \Psi^{pq} \mathfrak{F}_{p}\wedge \mathfrak{F}_{q}+\rho \Psi^{pq}Tr(t_pt_q),
\end{equation}
we note that only the symmetric part of the $\Psi$ field plays a role in the action, since the trace of the Lie algebra generators and the wedge product of a couple of two-forms are both symmetric in its indices. We have to draw attention to the fact that this symmetric behavior only works for a pure bosonic case, since, for the supersymmetric case, the field $\Psi$ inherits graded symmetry coming from the product of the FF wedge product \cite{Rosales-Quintero2016}. \\
Let us consider the equations of motion for $\rho$
\begin{equation} \label{eq: Constraint on Psi}
    \Psi^{pq} Tr(t_pt_q) =0,
\end{equation}
we have chosen  to impose first this constraint  on the field $\Psi$, before we consider to obtain the equation of motion coming from the variation of the $\Psi$ field, but since the equations of motion coming from the action are purely algebraic type then we should proceed first by considering the variation of $\Psi$ and then the constraint imposed by $\rho$ and the result is the same, so, since the order is not important and the two roads are equivalent, we chose a path that we consider is the more convenient.\\
Once we have constrained the $\Psi$ field through (\ref{eq: Constraint on Psi}) , which we will denote  it as $(\Psi)^{c}$, then the action is written as
\begin{equation}     \label{eq: Constrained action with indices }
    S=\int_{\mathcal{M}} \ (\Psi^{pq})^{c}\ \mathfrak{F}_{p}\wedge \mathfrak{F}_{q}
\end{equation}
hence the equation of motion for $(\Psi)^c$ implies
\begin{equation} \label{Eq: Simplicity constraint general}
    (\mathfrak{F}\wedge \mathfrak{F})^c=0.
\end{equation}
The kind of solutions that we are looking for, from this condition, are torsionless conformally flat Einstein manifolds \cite{Besse1987} defined by  \begin{equation} \label{eq: F prop to sigma}
\mathfrak{F}\propto e\wedge e\equiv \Sigma,
\end{equation}
where $e$ is the one-form tetrad field, letting the metric tensor $g$ appears  through $g=e\otimes e$, therefore endowing metric structure into the pure gauge theory, also, we define the Lie algebra valued two-form field $\Sigma$ as the wedge product of tetrad fields. The set equations (\ref{Eq: Simplicity constraint general})  which implies (\ref{eq: F prop to sigma}) are known under the name of simplicity constraints, they are algebraic type equations and they are known in the literature from BF formulations for gravity \cite{plebanski, Capovilla2001}. \\
The torsionless property of these manifolds can be seen  by applying the $G$ covariant derivative $D_{A}$ to (\ref{eq: F prop to sigma}) and by considering the Bianchi identity $D_{A}F=0$, where we obtain the zero torsion condition $D_{A}\Sigma=0$. This equation enforces  $A$ to be a function of the frame field $A=A(e)$. The constant of proportionality in (\ref{eq: F prop to sigma}) is related to the cosmological constant, $\Lambda$, due to dimensional consistency \cite{Torre1990}. \\
 In consequence, in an attempt to obtain (anti-) self-dual solutions, e.i., Einstein manifolds, it is necessary  to give suitable degrees of freedom to the $\Psi$ field, by taking an adequate  traceless condition given by the  Lagrange multiplier $\rho$. The case for the symmetry groups $SU(2)$ and $OSp(1\mid 2)$ can be checked in \cite{Rosales-Quintero2016}. Due to the fact that we are interested in a MacDowell-Mansouri type formulation and, its group of symmetry contains $SO(1,3)$ as subgroup, then here we will briefly present the $SO(1,3)$ case, since some of the tools for this subgroup will be necessary for the total group itself  (for full details for $SO(1,3)$ formulation check  \cite{Rosales-Quintero2020}).


\subsection{SO(1, 3) formulation}

As was described in the last section, the fundamental field is the connection one-form, in this case let us consider the gauge group as $SO(1,3)$, then the connection is characterized by   $A=A^{ab}t_{ab}$, where $A^{ab}$ is the spin connection, so the field strength is given by  $F=(dA^{ab}+A^{ac}\wedge A_{c}^{\ b})t_{ab}$, where $t_{ab}$ the $\mathfrak{so(1,3)}$ Lie algebra generators, satisfying
\begin{displaymath}
  [t_{ab},t_{cd}]=-\frac{1}{4}\ \eta_{ab,[c}^{\ \ \ \ \ \ [e}\eta_{d]}^{\ \ f]} t_{ef}
\end{displaymath}
so the CKf is given forthright as 
\begin{displaymath}
  \kappa_{abcd}=\frac{1}{I_{ad}} \ tr(t_{a}t_{b})=\frac{1}{2}\ \eta_{ab,cd}.
\end{displaymath}
However, in $\mathfrak{so(1, 3)}$, there is another invariant trace that can be specified, and it is given by
\begin{displaymath}
  \kappa_{abcd}^{1}=\frac{1}{I_{ad}^{1}} \ tr^{1}(t_{a}t_{b})=\frac{1}{I_{ad}^{1}} \ tr(*t_{a}t_{b})=\frac{1}{2}\ \epsilon_{abcd}
\end{displaymath}
where $\epsilon_{abcd}$ is the Levi-Civita tensor and $*$ is the Hodge dual acting on a pair of $\mathfrak{so(1,3)}$ Lie algebra valued fields, defined as $*\xi^{ab}=\frac{1}{2}\ \epsilon^{abcd}\xi_{cd}$. Once, we have introduced the two invariant traces that can be defined to construct gauge invariant forms in $SO(1,3)$, let us define the trace, ``Tr' as a linear combination of these forms
\begin{equation} \label{eq: General trace definition}
Tr(t_{ab}t_{cd})=a_{1} \ tr(t_{ab}t_{cd})+a_{2} \ tr(*t_{ab}t_{cd})=\frac{1}{2}\ \bigg( a_{1}\eta_{ab,cd}+a_{2}
\epsilon_{abcd} \bigg):=\Pi_{ab,cd},
\end{equation}
where $a_{1}$ and $a_{2}$ are  constants. We have to remark that the CKf, $\kappa$,  is used to associate vectors and covectors in the Lie algebra, and the linear combination of forms, $\Pi$,  is introduced  into  the definition of the action. Finally, we have chosen to  rename  ``Tr'' as $\Pi$ in order to associate it to a pseudoprojector, i.e., a map from Lie algebra bivectors to a Lie algebra bivectors such that $\Pi \cdot \Pi\neq\Pi$.\\
Hence, in this case, the action reads (\ref{eq: Rosales Action indices version})
\begin{equation}  \label{eq: Rosales action so(1,3)} 
S_{FF}[A, \Psi, \rho ]=\int_{\mathcal{M}} 
 \Psi^{abcd}\mathfrak{F}_{ab} \wedge \mathfrak{F}_{cd} + \rho \Psi^{abcd}\Pi_{abcd}
\end{equation}
where $\mathfrak{F}_{ab}=\Pi_{abcd}F^{cd}$.  As we have mentioned in the last section, the equation of motion coming from $\rho$  constraints the degrees of freedom for the field $\Psi$ in such a way that when we consider the equation of motion $\Psi^{c}$, we obtain, in (\ref{eq: Rosales action so(1,3)}), the CMPR simplicity constraint \cite{Capovilla2001} for the two-form field $\mathfrak{F}$, so the solution is
\begin{equation}   \label{eq: simplicity}
\mathfrak{F}^{ab}=\Pi^{abcd}F_{cd}=\Lambda(\alpha \Sigma^{ab}+\beta *\Sigma^{ab}) 
\end{equation}
where $\Lambda$ is the cosmological constant and $\Sigma^{ab}=e^{a}\wedge e^{b}$, $e$ being the tetrad one-form and $\alpha$, $\beta$ are constant dimensionless terms that satisfy $\alpha^{2} \neq \beta^{2}$
\begin{equation}   \label{eq: ratio a2/a1}
    \frac{a_{2}}{a_{1}}=\frac{\beta^2-\alpha^2}{4\alpha\beta}.
\end{equation}
The zero torsion condition, $T^{a}=0$ which implies $ A=A(e)$, can be obtained  when  we apply the $SO(1,3)$-covariant derivative on both sides in (\ref{eq: simplicity}), and by using the Bianchi identity, $DF=0$.\\
Since we are interested not in $\mathfrak{F}$ but $F$ alone, we have to apply the inverse of $\Pi$, which it is given by  
\begin{equation}
    \Pi^{-1}_{ab,cd}=\frac{1}{2}\ \frac{1}{a_{1}^2+a_{2}^2}\ (a_{1}\eta_{ab,cd}-a_{2}\epsilon_{abcd}),
\end{equation}
where $(\Pi^{-1}\Pi)_{ab,cd}=(\Pi\Pi^{-1})_{ab,cd}=\frac{1}{2}\ \eta_{ab,cd}$. Finally, the $\mathfrak{so(1,3)}$ field strength is written as function of only two free parameters $\{ \alpha, a_{1} \} $ or $\{ \beta, a_{2} \} $, as follows
\begin{equation} \label{eq: Einstein Manifolds}
 F_{ab}=\frac{\alpha}{a_{1}}\ \Lambda\Sigma_{ab}=\frac{\beta}{a_{2}}\ \Lambda\Sigma_{ab}.
\end{equation}
where the $*\Sigma$ sector was eliminated by means of the equation  $DT^{a}=F^{ab}\wedge e_{b}=0$, and together with (\ref{eq: ratio a2/a1}), they both imposed conditions on the values of $\alpha, \beta, a_{1}$ and $a_{2}$. A final remark: it is important to note that, without loss of generality, both  quotients $\alpha/a_{1}$ and $\beta/a_{2}$ are real valued \cite{Rosales-Quintero2020}.\\ 
As we can observe from (\ref{eq: Einstein Manifolds}), we obtain our main objective, torsionless Einstein manifolds, additionally these  are maximally symmetric spaces, i.e., they are  isotropic and homogeneous spaces. \\
Lastly,  the equation motion for $A$ gives
\begin{equation}    \label{eq: Integrability condition}
  D\Psi^{abcd}F_{cd}=0,
\end{equation}
implying that $\Psi^{abcd}F_{cd}$ is covariantly constant.

\section{SO(1, 4) formulations}
In this section, we will consider two different formulations for a  gauge theory of gravity where the gauge group $G \supset SO(1, 3)$. $G$ is usually taken in such a way that it depends on the sign of the cosmological constant $\Lambda$. Here, due  to phenomenological considerations,   $\Lambda$ is assumed positive so we will take  as our gauge group $SO(1,4)$ 
(but the case with negative cosmological constant, $SO(2, 3)$, can be analyzed straightforward). The connection $A$ is an $\mathfrak{so(1,4)}$-valued one-form on $\mathcal{M}$ and it is given by  $A=A^{\ AB}_{\mu}t_{AB} dx^{\mu}$, where $t_{AB}$ are the skew-symmetric  generators of the Lie algebra $\mathfrak{so(1,4)}$ fulfilling 
\begin{equation}
[t_{AB},t_{CD}]=f_{ABCD}^{\ \ \ \ \ \ \ \ \ EF} \ t_{EF}=\frac{1}{4}\
\eta_{AB,[C}^{\ \ \ \ \ \ \ \ [E }\eta_{D]}^{\ \ F]} t_{EF}
\end{equation}
then the CKf can be calculated forthright, $\kappa_{ABCD}=\frac{1}{2}\ \eta_{AB,CD}$. Now, since $\mathfrak{so(1,4)}$ is a reductive geometry then it could be written, as vector space, as $\mathfrak{so(1,4)}\cong \mathfrak{so(1,3)} \oplus \mathbb{R}^{1+3}$ \cite{Wise2010, Wise2009}. This decomposition  enables us to write the  indices as follows: $\{ AB\}=\{ ab,a4 \}$. Consequently, it is possible  to  make the following identification for the connection 
\begin{equation}     \label{eq: Decomposition Connection}
A^{AB}= \left(\begin{array}{cc}
    A^{ab}  & A^{ a4} \\
    A^{ 4b}  & 0
\end{array}\right)=
\left(\begin{array}{cc}
  A^{ ab}     & -\frac{1}{l}\ e^{a} \\
   \frac{1}{l}\ e^{b} & 0
\end{array}\right)
\end{equation}
where $A^{ab}$ and  $e^{a}$, as in the previous section,  are the spin connection and the tetrad field accordingly, and $l$  is known as the radio of the mathematical (Anti-)De Sitter space \cite{Blagojevic:2002du}, it has dimension of length necessary for unit requirements.  The field strength is given by
\begin{equation} \label{eq: Tensor field in so(1,4)}
F^{ AB}= \left(\begin{array}{cc}
    F^{ ab} & F^{a4} \\
    F^{ 4b}  & 0
\end{array}\right)=
\left(\begin{array}{cc}
    R^{ab}-\frac{1}{l^2}\ \Sigma^{ab}            & -\frac{1}{l}\ T^{a} \\
  \frac{1}{l}\ T^{b}                             & 0
\end{array}\right)
\end{equation}
where $T^{a}$ and $R^{ab}$  are the $SO(1,3)$ torsion and field strength. We have to note that in this section, we have made some notation changes for the field strength,  $R^{ab}$ is the $\mathfrak{so(1,3)}$ field strength and $F^{ab}$ is the $\mathfrak{so(1, 3)}$ part of the full $\mathfrak{so(1, 4)}$ field strength.\\
As was noted independently by MacDowell-Mansouri \cite{MacDowell1977}, and Stelle-West \cite{Stelle1979}, in order to construct a gauge theory  of gravity with nonvanishing cosmological constant that  reproduce the usual four-dimensional geometry of gravity that considers  the $\mathfrak{so(1,4)}$ valued  connection as the fundamental field, it is necessary to break the symmetry group by introducing some specific  constraints. In the next two subsections, we will briefly explain some important  details that we will be necessary for each symmetry breaking approach, MM and SW, with a view to apply these results  to obtain explicitly  the conformally flat Einstein manifolds.


\subsection{MacDowell-Mansouri Approach}
In this subsection, we will consider the approach given in \cite{Lopez-Dominguez2018}. In there,  the main idea is to consider as the Lagrangian density $\mathcal{L}=Tr \ F\wedge F$ where 
\begin{equation}
    Tr(t_{AB}t_{CD})=a\kappa_{ABCD}-a_{1}\kappa^{1}_{ABCD}+a_{2}\kappa^{2}_{ABCD}
\end{equation}
where $a, a_{1}$ and $a_{2}$ are constant terms and
\begin{equation}   \label{eq: Cartan killing form projected so(3,1) 1}
\kappa^{(1)}_{ABCD}=
\left(
\begin{array}{cc}
 \kappa^{(1)}_{abcd} & \kappa^{(1)}_{abc4} \\
 \kappa^{(1)}_{a4cd}  & \kappa^{(1)}_{a4c4}
\end{array}\right)=
-\frac{1}{2}\left(
\begin{array}{cc}
 \eta_{ab,cd} & 0 \\
 0  & 0
\end{array}\right),
\end{equation}
and
\begin{equation}   \label{eq: Cartan Killing form projected so(3,1) 2}
\kappa_{ABCD}^{(2)}=
\left(
\begin{array}{cc}
 \kappa^{(2)}_{abcd} & \kappa^{(2)}_{abc4} \\
 \kappa^{(2)}_{a4cd}  & \kappa^{(2)}_{a4c4}
\end{array}\right)=\frac{1}{2}
\left(\begin{array}{cc}
\epsilon_{abcd} & 0 \\
 0 & 0
\end{array}\right),
\end{equation}
As we have told, MM formulation is based in an symmetry breaking scheme, and it is done by considering the extra terms in the definition of $Tr$, i.e.,  $\kappa^{1}$ and  $\kappa^{2}$ are introduced in order to break the symmetry from $SO(1, 4)$ down to $SO(1,3)$ and they are built by considering the fundamental representation for the Lie algebra generators $t_{AB}$ \cite{Lopez-Dominguez2018}. \\
Then $Tr$ can be written as
\begin{equation}  \label{eq: CKf two sectors}
Tr(t_{AB}t_{CD})=\frac{a}{2}\ \eta_{AB,CD}
+\left(
\begin{array}{cc}
 \Pi_{abcd} & 0 \\
 0  & 0
\end{array}\right),
\end{equation}
where $\Pi$ is defined in (\ref{eq: General trace definition}). As we may observe, the last equation has two different sectors, one is related to the full $\mathfrak{so(1,4)}$ Lie algebra, and the second one is related to the symmetry breaking sector that let us project the full Lie algebra valued fields to the $\mathfrak{so(1,3)}$ Lie algebra valued fields. \\
With all the ingredients at hand, the action gives:
\begin{displaymath}  
S_{GMM}\left[ A\right]= \int_{\mathcal{M}}  {\rm Tr} \ F\wedge F=
\int_{\mathcal{M}}\ F^{AB}\wedge F^{CD}\
Tr(t_{AB}t_{CD})
\end{displaymath}
\begin{eqnarray}      \label{eq: GMM-Rosales}
&=&a_{2}\int_{\mathcal{M}}\frac{a+a_{1}}{a_{2}}R^{ab}\wedge
R_{ab}+\frac{1}{2} R^{ab}\wedge R^{cd}
\epsilon_{abcd}+\frac{2a\Lambda}{3a_{2}} d(e^{a}\wedge T_{b}) \nonumber\\
&~ &-\frac{a_{2}\Lambda}{3}\int_{\mathcal{M}}\left[ \epsilon_{abcd}\ R^{ab}\wedge \Sigma^{cd}+\frac{2a_{1}}{a_{2}} R^{ab}\wedge
\Sigma_{ab}  -
\frac{\Lambda}{6}\ \epsilon_{abcd}\ \Sigma^{ab} \wedge \Sigma^{cd}\right],
\end{eqnarray}
where  the cosmological constant is defined as $\Lambda=3l ^{-2}$ and the Immirzi parameter can be associated with the quotient
$a_{2}/a_{1}\sim\gamma$. The first
three terms in the action correspond to the topological sector, they  are  second Chern class, the Euler class and the Nieh–Yan Class and the last three correspond to the dynamical sector, and they are the Palatini term, the Holst term and the cosmological constant term, correspondingly.\\
Now, since we are interested in  conformally flat Einstein manifolds in a MM formulation,  let us consider the action (\ref{eq: Rosales Action}) where $A$ and $\Psi$ are   $\mathfrak{so(1,4)}$-valued fields.\\
\begin{equation} \label{eq: FF action in MM}
    S_{FF} [A,\Psi, \rho]=\int_{\mathcal{M}} \ \Psi^{ABCD} F^{EF} Tr(t_{AB}t_{EF}) \wedge F^{GH}Tr(T_{CD}t_{GH})+\rho \Psi^{ABCD}Tr(t_{AB}t_{CD}) 
\end{equation}
but before we proceed to obtain the Einstein Manifolds, it is important to observe that the action (\ref{eq: GMM-Rosales}) actually is contained in (\ref{eq: FF action in MM}). In fact, if we choose $\Psi^{ABCD} \rightarrow \frac{1}{2} \eta^{AB, CD} $ and $\rho \rightarrow 0$ in (\ref{eq: FF action in MM}) then the action (\ref{eq: GMM-Rosales}) is  achieved, modulo  a redefinition of the constants $a, a_{1}$ and $a_{2}$. The latter condition is not completely necessary since if we do not consider a vanishing $\rho$ then it imposes a relation on $a$ and $a_{1}$ that do not affect the general functional form of the resulting action. The former condition is more interesting since it shows that  (\ref{eq: GMM-Rosales})  can be seen as a base state of the more general action (\ref{eq: FF action in MM}).  \\ 
Now, let us go back to our initial objective, that is, describe the conditions under, from (\ref{eq: Rosales Action}), we obtain Einstein manifolds in a MM approach. As it was pointed out in section 2,  we have to be careful in considering the constraints related to the field $\Psi$ through the Lagrange multiplier  $\rho$, and on the other hand, since  MM formulation involves a symmetry breaking scheme, then we propose that the trace term involves only $\kappa^{1}$ and $\kappa^{2}$:
\begin{equation}  \label{eq: CKf two sector proyected}
    Tr(t_{AB}t_{CD})= \left(
\begin{array}{cc}
 \Pi_{abcd} & 0 \\
 0  & 0
\end{array}\right),
\end{equation}
so we observe that we are considering $a=0$ in (\ref{eq: CKf two sectors}). Also, this choice let us project our action to the  $\mathfrak{so(1,3)}$-sector. Then action (\ref{eq: FF action in MM}) reads
\begin{equation}
S_{FF}=\int_{\mathcal{M}} \ \Psi^{abcd} F^{ef} \Pi_{abef} \wedge F^{gh}\Pi_{cdgh}+\rho \Psi^{abcd}\Pi_{abcd}. 
\end{equation}
The last equation is very similar to (\ref{eq: Rosales action so(1,3)}), but they both differ in the functional form of the strength  tensor, since for this case, $F^{ab}$ is the $\mathfrak{so(1,3)}$ projection of the strength  tensor  defined in (\ref{eq: Tensor field in so(1,4)}). Now, let us proceed as in the pure $SO(1,3 )$ case (last section). First, by taking into account the equation of motion from the variation for $ \rho$, the solution for the  equation of motion for the constraint $\Psi$ implies
\begin{equation} \label{eq: Simplicity in MM }
 \Pi_{abcd} \bigg( R^{cd}-\frac{1}{l^2}\Sigma^{cd} \bigg)= \Lambda (\alpha \Sigma_{ab}+\beta*\Sigma_{ab}).
\end{equation}
where (\ref{eq: ratio a2/a1}) is still valid here. Then, by applying the $SO(1, 3)$-covariant derivative on both sides of the last equation, we obtain the zero torsion condition. Moreover, by applying a second covariant derivative and taking into account (\ref{eq: ratio a2/a1}), we finally obtain
\begin{equation}  \label{eq: Einstein manifolds MM}
    R^{ab}=\bigg[  \frac{\alpha}{a_{1}}+\frac{1}{3}    \bigg] \Lambda \Sigma^{ab}= \bigg[  \frac{\beta}{a_{2}}+\frac{1}{3}    \bigg] \Lambda \Sigma^{ab}, 
\end{equation}
then we obtain Einstein torsionless manifolds parameterized by two numbers as in the pure $SO(1, 3)$ case. Thus, in this MM formulation, we obtain a $1/3$ factor that can be absorbed by a redefinition of the parameters appearing in (\ref{eq: Einstein manifolds MM}).   

\subsection{Stelle-West approach}
SW approach, as well as MM, consider a formulation based on the gauge group $SO(1, 4)$ (or $(SO(2, 3)$ correspondingly), where the symmetry breaking mechanism is given not in the definition of $Tr$, instead it is introduced by an auxiliary   $\mathfrak{so(1, 4)}$-vector field $v$ on  $\mathcal{M}$, that has to be chosen in such a way that the symmetry is broken down to $SO(1, 3)$. This vector field is related with the radius of the (Anti) De Sitter space \cite{Stelle1979,Stelle1980}.\\
In this subsection, we will consider the  action proposed in \cite{DIaz-Saldana2021}. It can be written as follows
\begin{equation}  \label{eq: GSW-Rosales}
    I_{GSW} [A,v]=\int_{\mathcal{M}} \frac{1}{4}\ (\theta \eta^{AB,CD}+\epsilon^{ABCDE}v_{E})(\theta \eta_{AB,FG}+\epsilon_{ABFGH}v^{H}) F_{CD} \wedge F^{FG}+\rho (v^{E}v_{E}-l^{-2}),
\end{equation}
then, in order  to break the symmetry it is  necessary  to consider the  condition given by $\rho$ and impose that  $v^{E}=(0,0,0,0, 1/l)$. Then, when we substitute back into the action, we obtain an equivalent action as (\ref{eq: GMM-Rosales}), i.e., a theory of  GR with non zero cosmological constant and topological terms where $\theta=\gamma/l$ \cite{DIaz-Saldana2021}.\\
Before we proceed to our case of interest, let us make the following definitions
\begin{eqnarray}
\frac{1}{4}\ (\theta \eta^{AB,CD}+\epsilon^{ABCDE}v_{E})(\theta \eta_{AB,FG}+\epsilon_{ABFGH}v^{H}) & = & \Psi^{CD}_{\ \ \ \ FG}\\
Tr(t_{AB}t_{CD}) &=& \frac{1}{2}\ \eta_{AB,CD}\\
v^{E}v_{E}-l^{-2} & \approx &  Tr(\Psi)=\Psi^{AB}_{\ \ \ \ AB}.
\end{eqnarray}
Then (\ref{eq: GSW-Rosales}) is rewritten as 
\begin{equation}  \label{eq: I_GSW almost equal }
    I_{GSW} [A,v] \approx \int_{\mathcal{M}} Tr(\Psi(F)\wedge F)+\rho Tr(\Psi),
\end{equation}
where we have to note how now it is defined  $Tr$, in this case is just the usual CKf. Also, (\ref{eq: I_GSW almost equal }) is almost equivalent to (\ref{eq: Rosales Action}) and  the difference  from the original formulation \cite{DIaz-Saldana2021} is given by the constraint term, in our case, we have  $Tr(\Psi)=-6[v^{E}v_{E}-l^{-2}(5\gamma^{2}/3)]$, i.e., there is an extra term, $(5\gamma^{2}/3)$ that appears into the constraint, however, we can solve this situation by using two different paths, one of them is related to the limit when $l$ is large, i.e., when the cosmological constant $\Lambda$ is small (as the actual cosmological models suggest) then both formulations coincide. The second one consists in considering the following reescaling into the action:  $v \rightarrow \overline{v}=\gamma \sqrt{5/3}\ v $ and  $  \theta \rightarrow\overline{\theta}=\gamma \sqrt{5/3}\ \theta $ then (\ref{eq: I_GSW almost equal }) becomes an equality.\\
Then for the rest of this subsection, without lack of generality, let us consider that (\ref{eq: I_GSW almost equal }) is an equality. Therefore this action reflects the SW formulation once we have defined the functional form of $\Psi$ as function of $\theta$ and $v$ together to the symmetry breaking condition for $v$. But let us forget the functional structure for  $\Psi$ and let us consider only the equation (\ref{eq: I_GSW almost equal }) as our starting point, this is essentially equation (\ref{eq: Rosales Action}). As our main objective is to obtain Einstein manifolds, and, as opposed as MM formulation in the last subsection, the trace term in the action, $Tr$, is just the CKf, then we can not consider the symmetry breaking in this way, so in this case,  we must break the symmetry by considering certain condition over the $Tr(\Psi)$, that let us project this tensor field down to $SO(1, 3)$ then let us consider only for the constraint term
\begin{equation}
    Tr(\Psi) \rightarrow \Pi^{abcd}\Psi_{abcd},
\end{equation}
this is the symmetry breaking condition \textit{a l\'a} SW for Einstein Manifolds. Then the action can be written as
\begin{equation} \label{eq: I_GSW in SW}
    I_{GSW}=\int_{\mathcal{M}} \Big[ \Psi^{abcd} F_{ab} \wedge F_{cd}+\rho \Psi^{abcd} \Pi_{abcd} \Big] +2\Psi^{a4cd} F_{a4} \wedge F_{cd} +F^{abc4} F_{ab}\wedge F_{c4}.
\end{equation}
We can identify, for the terms inside the square brackets, similar full $\mathfrak{so(1, 3)}$ valued relations found in the last section for MM and $SO(1, 3)$ cases. Then, by considering the equations of motion, we obtain the functional form for $R$
\begin{equation}
    R_{ab}=\Lambda \bigg[ \bigg(\alpha +\frac{1}{3} \bigg) \Sigma^{ab}+\beta*\Sigma^{ab} \bigg],
\end{equation}
so if we  proceed the similar way as in the last section, by applying a $SO(1, 3)$ covariant derivative, we obtain the torsionless condition, in this case can be written as $F^{a4}=0=F^{4a}$, then the rest terms in the action (\ref{eq: I_GSW in SW}) vanishes identically, so they do not offer any dynamical contribution. Now, by applying a second covariant derivative and taking into account (\ref{eq: ratio a2/a1}), which it is still valid here, we get $\beta=0$, then the spaces modeled by this action are  given by
\begin{equation}
    R_{ab}=\Lambda \bigg[ \alpha +\frac{1}{3} \bigg] \Sigma^{ab},
\end{equation}
then in this formulation Einstein Manifolds are solutions parametrized by just one term. But in this case we must be very careful, since, once that we have found that $\beta$ vanishes then $\alpha$ can not vanish too.

\section{Conclusions and Outlooks}
In this work we have considered a constrained pure connection formulation, that we have named it as $\Psi(F)$ formulation, which is based on the works  given by Smolin \cite{smolin2002quantum} and Torre \cite{Torre1990}, but taking into account the modifications given by the author himself in a pair of previous papers \cite{Rosales-Quintero2016, Rosales-Quintero2020} for the gauge groups:  $SO(1,3)$, complex $SU(2)$ and the supergroup $OSp(1, 2)$. In this case,  we took as a gauge group $SO(1, 4)$ (but it works as well for $SO(2, 3)$), but as the works given, on one side, by  MacDowell and Mansouri \cite{MacDowell1977}, and on the other side, by Stelle and West \cite{Stelle1979, Stelle1980}, has shown, is that in order to obtain GR plus cosmological term and topological invariant terms, it is necessary to break  the symmetry down to $SO(1, 3)$ by means of constraints terms. The form of the constraint terms is inspired by the works \cite{Lopez-Dominguez2018, DIaz-Saldana2021} for MM formulation and SW formulation. Once the constraint terms are introduced into the action (\ref{eq: Rosales Action}), the equations of motion led us to obtain, as solution of the equations of motion, torsionless Einstein manifolds. Also, we observe that the MM formulation and SW formulation can be written as $\Psi(F)$ formulation, this suggest that this last formulation could be more general formulation than the first two. Also, as the author thinks, this is not a trivial observation, and could led,  through different constraints,  to GR, by much simpler way or/and where the constraint term has a different geometric meaning. The last subject is studied by the author and the results  will be publish elsewhere.\\
Also, it is interesting to note that if we consider a SW approach for the action (\ref{eq: GSW-Rosales}) then the Immirzi parameter appears in the radio of the mathematical (Anti)-De Sitter  space. This could show that the value of the cosmological constant is affected by the  appearance of Immirzi parameter, at last in this model. It could be interesting to compare both models (with or without Immirzi parameter) in order to show the differences at the quantum level, when the Immirzi parameter plays an important role. This  work is in progress and it will be presented somewhere else.\\      
Canonical analysis for $\Psi(F)$ formulation for the gauge group $SO(1, 4)$ (or $SO(2, 3)$) will be important  as a first look look for a quantization program for this action.


\section*{Acknowledgements}
J. E. Rosales-Quintero is supported by  "Est\'imulos a la investigaci\'on para doctoras y doctores 2020 del  Consejo de Ciencia y Tecnolog\'ia del Estado de Puebla", he also acknowledges support from Benem\'erita Universidad Aut\'onoma de Puebla  for granting the necessary conditions to carry out this research work.


\bibliographystyle{unsrt}  
\bibliography{references}  

\end{document}